
%
%
\input harvmac

\Title{UCLA/92/TEP/46}{Current-Current Singularities
under External Gauge Fields$^*$}
\footnote{}{*This work was supported in part by the U.S. Department
of Energy, under Contract DE-AT03-88ER 40384 Mod A006 Task C.}

\centerline{Hidenori SONODA}
\bigskip\centerline{\it Department of Physics,
UCLA, Los Angeles, CA 90024--1547, USA}

\vskip 1in
We study the
operator product expansion of two non-interacting
chiral
currents in the presence of external
gauge fields in four dimensional euclidean space.
We obtain the operator singularity in terms of
the beta function of the free energy density and a connection
on the space of external
gauge fields,
by imposing the consistency between
the variational formula, introduced previously
by the author, and both the renormalization group
equations and the equation of motion for the currents.
As a byproduct, we derive a euclidean version of the
anomalous commutator of two currents in an appendix.

\Date{December 1992}

\def\ave#1{\left\langle #1 \right\rangle}
\def\O{{\cal O}}
\def\tr{{\rm tr~}}
\def\Sp{{\rm Sp~}}
\def\F{{\cal F}}
\def\gone{g_{\bf 1}}
\def\bone{\beta_{\bf 1}}
\def\e{{\rm e}}
\def\p{\partial}
\def\ep{\epsilon}
\def\d{\delta}
\def\dl{{d \over dl}}
\def\one{{\bf 1}}
\def\r{\hat{r}}
\def\tE{\tilde{E}}
\def\tF{\tilde{F}}
\def\tG{\tilde{G}}
\def\tH{\tilde{H}}
\def\Om{\Omega}
\def\C{{\cal C}}
\def\psibar{\overline{\psi}}
\def\proj{{1 - \gamma_5 \over 2}}
\def\gm{\gamma}

\newsec{Introduction}

In thermodynamics the concept of {\it conjugacy} plays important
roles.  Intensive variables have conjugate extensive variables;
for example a constant external magnetic field $H$ has
total magnetization $M$ as its conjugate
variable.  We can formulate the conjugacy relation
quantitatively by using a Gibbs free energy $G(H)$:
the correlation of $n$-number of $M$'s is equal to
the n-th order derivative of $G(H)$ with respect to $H$.
Namely, we find
\eqn\ethermo{\langle \underbrace{M ... M}_n \rangle^c =
(-)^{n-1} {\partial^n G(H) \over \partial H^n} ,}
where $^c$ denotes the connected part.
For $n=2$ we obtain the magnetic susceptibility as
the connected two-point correlation of magnetization.  This
is one example of the famous fluctuation-dissipation theorem.
\ref\rkubo{R. Kubo, J. Phys. Soc. Japan {\bf 12}(1957)570}

In renormalized field theories the singularities of operator
products at short distances
complicate the generalization of
eq.~\ethermo.  We have examined this question in detail
in two previous
papers \ref\rqcd{H. Sonoda,
Nucl. Phys. {\bf B383}(1992)173}
and \ref\rphi{H. Sonoda,
``Operator coefficients for composite operators
in the $(\phi^4)_4$ theory,'' to appear in Nucl. Phys. B};
we regard renormalized parameters $g^i$, like $\lambda$ in the
$\phi^4$ theory, as constant intrinsic variables and introduce
spatial integrals of renormalized composite operators
$\O_i$, like $\phi^4 (r)$,
as their conjugates.  We call the generalization of eq.~\ethermo\
as variational formulas.  We have obtained the singularity
of the product of $\O_i$ and an arbitrary composite operator
$\Phi_a$ in terms of beta functions, anomalous dimensions,
and finite counterterms $c_{ia}^b$ that appear in the variational
formulas.  We have also obtained the geometrical interpretation
of $c_{ia}^b$ as a connection for the vector bundle of composite
operators over the theory space.  In ref. \rphi\ we have
applied the variational formula to derive perturbation theory
for operator product coefficients.

The purpose of the present paper is to generalize the work
of refs.~\rqcd, \rphi\ from constant external fields
to spatially dependent external fields.  In particular we examine
non-interacting chiral $SU(N)$ currents $J_\mu (r)$
under external gauge fields $A_\mu (r)$ in four dimensional
euclidean space.

The paper is organized as follows.  In sect.~2 we will discuss
the renormalization group (RG) equations for the free energy
density and currents.  In sect.~3 we will briefly introduce
variational formulas.  In sect.~4 we will derive relations among
the coefficients of the operator product of two currents using
the current equation of motion.  In sect.~5 we will study the
structure of finite counterterms that appear in the variational
formulas.
The finite counterterms will be interpreted as a connection over
the space of external gauge fields.
In sect.~6 we will determine the current-current operator
singularities
in terms of the connection.
In sect.~7 we will determine some of the operator
singularities explicitly for the chiral fermion in the
fundamental representation of $SU(N)$.
Finally we conclude the paper
in sect.~8.  A large amount of calculations accompany our study;
we will give mainly results but only sketch the derivations,
in fear that excessive exposure of equations will spoil the
flow of the paper.  In an appendix we will derive the euclidean
version
of the anomalous commutator of two currents.

The current-current singularities
have been discussed first in
ref.~\ref\rwilson{K. Wilson, Phys. Rev. {\bf 179}(1969)1499}.
This reference is the primary inspiration for the present work.
For the case of a $U(1)$ external gauge field we can find pioneering
(and classic) works in refs. \ref\rjackiw{R. Jackiw, in
{\it Current Algebra and Its Applications} (Princeton, 1970);
S. Adler, in {\it Brandeis 1970 Lectures},
eds. S. Deser et al. (MIT Press, 1971)}.

\newsec{Renormalization group equations}

In four dimensional euclidean space we consider a chiral
$SU(N)$ current $J_\mu \equiv i T^a J_\mu^a$ and an
external field $A_\mu \equiv - i T^a A_\mu^a$, where
we normalize the generators $T^a$ in the fundamental
representation of $SU(N)$ by
\eqn\enorm{ \tr T^a T^b = \delta^{ab} .}
We assume that the current is a bilinear of
non-interacting chiral
fermion fields in some representation of $SU(N)$,
not necessarily the fundamental representation.
In the absence of $A_\mu$ the current
is free and conserved:
\eqn\econs{ \partial_\mu J_\mu = 0 \quad {\rm for}\quad A_\mu = 0 .}

Let $\F [\gone, A]$ be the total free energy:
\eqn\eF{\F [\gone, A] = \int d^4r~ \gone (r) + F[A] ,}
where $\gone (r)$ is a space-dependent cosmological
constant, and $F[A]$ gives the dependence on the external
gauge field.  Under the renormalization group (RG)
transformation by scale
$\e^l$\footnote{$^\dagger$}
{We adopt a convention in which the renormalization
scale is fixed under the RG.}, the total free energy, being physical,
is invariant, while
$\gone (r)$ and $A_\mu(r)$ change according to the following
RG equations:
\eqn\eRG{\eqalign{
{d \gone (r) \over dl} &= 4 \gone (r) + \bone (A(r)) \cr
{d A_\mu(r) \over dl} &= A_\mu(r) .\cr}}
The beta function $\bone (A(r))$ must have scale dimension
4, dependent only on $A_\mu(r)$ and its spatial derivatives at $r$.
The field $A_\mu (r)$, with scale dimension 1, has nothing local to
mix with.

We can write the most general local form of the beta function:
\eqn\ebone{\eqalign{\bone (A) &=
b_1 ~\tr \p_\mu A_\nu \p_\mu A_\nu +
b_2 ~\tr \p_\mu A_\mu \p_\nu A_\nu \cr
&
+ b_3 ~\tr \p_\mu A_\nu A_\mu A_\nu +
b_4 ~\tr \p_\mu A_\nu A_\nu A_\mu \cr
&
+ b_5 ~\tr A_\mu A_\mu A_\nu A_\nu +
b_6 ~\tr A_\mu A_\nu A_\mu A_\nu \cr
&
+ b_7 ~\tr A_\mu A_\mu ~\tr A_\nu A_\nu +
b_8 ~\tr A_\mu A_\nu  ~\tr A_\mu A_\nu .\cr}}
We have neglected total derivatives; we can even neglect
the topological term
\eqn\etopo{\partial_\mu \big(
\ep_{\mu\nu\alpha\beta} \tr ( A_\nu \p_\alpha A_\beta +
{2 \over 3}~ A_\nu A_\alpha A_\beta ) \big) ,}
since it requires a discrete normalization
and cannot appear as a beta function of the free energy density.

The precise definition of the current $J_\mu (r)$ as a conjugate
to $A_\mu (r)$ can be given only in terms of the variational
formula that we introduce in the next section.  But we can
determine the RG equation of the current
just by knowing its expectation value.  One particular example
of the variational formula gives the dependence of the total
free energy on the external gauge field:
\eqn\evev{\d \F[\gone,A] = \d F[A]
= \int d^4 r~ \tr \d A_\mu (r) \ave{J_\mu (r)}_A .}
Then, the RG invariance of $\F$ and the above equation imply
\eqn\eRGvev{\dl \ave{J_\mu (r)}_A = 3 \ave{J_\mu (r)}_A
+ j_\mu (A(r)) ,}
where
\eqn\ej{\eqalign{\tr (- i T^a) j_\mu (A(r)) &\equiv - {\d \over \d
A_\mu^a (r)}
\int d^4 r' \bone(A(r')) \cr
& = \tr (- i T^a) \Big[ 2 b_1 \p^2 A_\mu + 2 b_2 \p_\mu \p_\nu A_\nu
\cr
&\quad + b_3 \left( \p_\nu (A_\nu A_\mu) - A_\nu \p_\mu A_\nu -
\p_\nu
A_\mu A_\nu \right) \cr
&\quad + b_4 \left( \p_\nu (A_\mu A_\nu) - A_\nu \p_\nu A_\mu -
\p_\mu A_\nu A_\nu \right) \cr
&\quad - 2 b_5 (A_\mu A_\nu A_\nu + A_\nu A_\nu A_\mu )
- 4 b_6 A_\nu A_\mu A_\nu \cr
&\quad - 4 b_7 A_\mu \tr (A_\nu A_\nu) -
4 b_8 A_\nu \tr (A_\nu A_\mu) \Big] .\cr}}
Since the current $J_\mu$ has only the identity operator to mix with,
the above RG equation is valid as an operator equation:
\eqn\eRGj{\dl J_\mu (r) = 3 J_\mu (r) + j_\mu (A(r)) \one .}

Before closing this section, we must notice that the
current $J_\mu$ has ambiguity.  We can change
the total free energy $\F[\gone,A]$ by
adding an integral of a local function of $A_\mu (r)$
to $F[A]$:
\eqn\echangeF{F[A] \to F'[A] = F[A] + \int d^4 r ~f(A(r)) .}
Here $f(A(r))$ is an arbitrary local function
of $A(r)$ with scale dimension 4:
\eqn\ef{\eqalign{f(A(r)) &=
a_1 ~\tr \p_\mu A_\nu \p_\mu A_\nu +
a_2 ~\tr \p_\mu A_\mu \p_\nu A_\nu \cr
&
+ a_3 ~\tr \p_\mu A_\nu A_\mu A_\nu +
a_4 ~\tr \p_\mu A_\nu A_\nu A_\mu \cr
&
+ a_5 ~\tr A_\mu A_\mu A_\nu A_\nu +
a_6 ~\tr A_\mu A_\nu A_\mu A_\nu \cr
&
+ a_7 ~\tr A_\mu A_\mu ~\tr A_\nu A_\nu +
a_8 ~\tr A_\mu A_\nu  ~\tr A_\mu A_\nu .\cr}}
According to \evev, this shifts the current operator
by a c-number:
\eqn\eredefJ{\tr (- i T^a) J'_\mu (r) = \tr (-i T^a) J_\mu (r) +
{\d \over \d A_\mu^a (r)} \int d^4 r' f(A(r'))  .}

\newsec{Variational formulas}

Let us consider a correlation of $n$ composite operators
$\Phi_1, ..., \Phi_n$ under an external gauge field $A_\mu$,
$\ave{\Phi_1 ... \Phi_n}_A$.  We would like a
variational formula that gives
the change of the correlation under an infinitesimal
change of $A_\mu$.
In refs. \rqcd, \rphi\ we have studied variational formulas
in the case of constant external fields.  By generalizing them
we obtain the desired formula in the following form:
\eqn\evar{\eqalign{&- \d \ave{\Phi_1 ... \Phi_n}_A \cr
&\quad = \lim_{\ep \to 0} \Bigg[ \int_{|r-r_i| \ge \ep} d^4 r~
\d A_\mu^a (r) \ave{(J_\mu^a (r) - \ave{J_\mu^a (r)}_A) \Phi_1 (r_1)
 ... \Phi_n (r_n)}_A \cr
&\quad\quad + \sum_{i=1}^n \Big( \d c_i^j (A(r_i)) - \int_{1 \ge |r|
\ge \ep}
{d^4 r \over 2 \pi^2}
\big[ \d A_\mu^a (r_i + r) C_{\mu i}^{a j} (r;A(r_i)) \big]_{div}
\Big) \cr
&\qquad\qquad\qquad \times
\ave{\Phi_1 (r_1) ... \Phi_j (r_i) ... \Phi_n (r_n)}_A
\Bigg] .\cr}}
Here $\d c_i^j (r_i;A)$ is a finite counterterm, $C_{\mu i}^{a j}
(r;A(0))$
gives the singular part of the operator product expansion (OPE)
\rwilson\
\eqn\esing{J_\mu^a (r) \Phi_i (0) = {1 \over 2 \pi^2}
C_{\mu i}^{a j} (r;A(0)) \Phi_j (0)
+ o \left( {1 \over r^4} \right) ,}
and $[...]_{div}$ denotes the singular part
(i.e., unintegrable part) with respect to $r$.
In the rest of this paper we will only study the current itself
as $\Phi$'s and will give more detailed discussions of
the finite counterterm and the OPE coefficient.

\newsec{Current-current singularities}

In the previous section we have introduced a variational formula.
But we still need to quantify the finite counterterm and the OPE
coefficient to give a precise meaning to the formula.

The relevant OPE is
\eqn\ejj{J_\mu^a (r) J_\nu^b (0) = {1 \over 2 \pi^2}
C_{\mu \nu}^{ab,\one} (r;A(0))
\one + o \left({1 \over r^4}\right) .}
The current $J_\mu$ has scale dimension 3; we obtain
only the identity operator as the singular part by
assuming the absence of operators of scale dimension 2 or
less other than the identity operator.

Now, by using the rotational invariance, global $SU(N)$ symmetry,
and the Bose symmetry

\eqn\ebose{J_\mu^a (r) J_\nu^b (0) = J_\nu^b (0) J_\mu^a (r) ,}
we can write down the most general form of the coefficient
function $C_{\mu \nu}^{ab,\one} (r;A(0))$:
\eqn\egeneral{\eqalign{& C_{\mu \nu}^{ab,\one} (r;A(0)) \cr
&\quad =
{C \over r^6} ~\delta^{ab} (\d_{\mu\nu} - 2 \r_\mu \r_\nu)
+ {1 \over r^5} ~D_{\mu\nu}^{ab} (r;A(0)) \cr
&\quad\quad +{1 \over r^4} ~\big(
E_{\mu\nu}^{ab} (r;A(0)) + F_{\mu\nu}^{ab} (r;A(0))
+ G_{\mu\nu}^{ab} (r;A(0)) +
H_{\mu\nu}^{ab} (r;A(0)) \big) ,\cr}}
where (we denote $\r_\mu \equiv r_\mu/|r|$)
\eqn\eD{\eqalign{D_{\mu\nu}^{ab} (r;A(0)) &\equiv
D_1 \tr \left( T^a T^b \r_\mu (A_\nu (0)
+ {r_\alpha \over 2} \p_\alpha A_\nu (0)) - (a \leftrightarrow b,
\mu \leftrightarrow \nu) \right) \cr
&+ D_2 \tr \left( T^a T^b \r_\nu (A_\mu (0) + {r_\alpha \over 2}
\p_\alpha
A_\mu (0)) - (a \leftrightarrow b,
\mu \leftrightarrow \nu) \right) \cr
&+ (D_3 \d_{\mu\nu} + D_4 \r_\mu \r_\nu) \tr [T^a, T^b]
\r_\alpha \left(A_\alpha (0) + {r_\beta \over 2} \p_\beta A_\alpha
(0)
\right) \cr
&+ D_5 \ep_{\mu\nu\alpha\beta} \tr \{T^a,T^b\} \r_\alpha \left(
A_\beta (0) + {r_\gamma \over 2} \p_\gamma A_\beta (0) \right) \cr}}
\eqn\eF{\eqalign{F_{\mu\nu}^{ab} (r;A(0)) &\equiv
F_1 \tr (T^a T^b \p_\mu A_\nu (0) +
(a \leftrightarrow b, \mu \leftrightarrow \nu) ) \cr
&+ F_2 \tr (T^a T^b \p_\nu A_\mu (0) +
(a \leftrightarrow b, \mu \leftrightarrow \nu) ) \cr
&+ (F_3 \d_{\mu\nu} + F_4 \r_\mu \r_\nu)
\tr \{T^a,T^b\} \p_\alpha A_\alpha (0) \cr
&+ (F_5 \d_{\mu\nu} + F_6 \r_\mu \r_\nu) \r_\alpha \r_\beta
\tr \{T^a,T^b\} \p_\alpha A_\alpha (0) \cr
&+ F_7 \left( \r_\mu \tr T^a T^b \r_\alpha \p_\alpha A_\nu (0) +
(a \leftrightarrow b, \mu \leftrightarrow \nu) \right) \cr
&+ F_8 \left( \r_\mu \tr T^a T^b \r_\alpha \p_\nu A_\alpha (0) +
(a \leftrightarrow b, \mu \leftrightarrow \nu) \right) \cr
&+ F_9 \left( \r_\nu \tr T^a T^b \r_\alpha \p_\alpha A_\mu (0) +
(a \leftrightarrow b, \mu \leftrightarrow \nu) \right) \cr
&+ F_{10} \left( \r_\nu \tr T^a T^b \r_\alpha \p_\mu A_\alpha (0) +
(a \leftrightarrow b, \mu \leftrightarrow \nu) \right) \cr}}
\eqn\eH{\eqalign{H_{\mu\nu}^{ab} (r;A(0)) &\equiv
H_1 \ep_{\mu\nu\alpha\beta} \tr [T^a,T^b] \p_\alpha A_\beta (0) \cr
&+ H_2 \left( \r_\mu \ep_{\nu\alpha\beta\gamma} \r_\alpha \tr
T^a T^b \p_\beta A_\gamma (0) +
(a \leftrightarrow b, \mu \leftrightarrow \nu)  \right) \cr
&+ H_3 \left( \r_\mu \ep_{\nu\alpha\beta\gamma} \r_\alpha \tr
T^b T^a \p_\beta A_\gamma (0) +
(a \leftrightarrow b, \mu \leftrightarrow \nu)  \right) .\cr}}
We give $E(r;A(0))$ and $G(r;A(0))$ in Appendix A; both are
quadratic in $A(0)$, and $G$ contains one $\ep$ tensor, while
$E$ contains none.

As it is, the coefficient function is hopelessly complicated.  We can
simplify the function tremendously by imposing the current equation
of motion:
\eqn\emotion{\p_\mu J_\mu^a + k_1 f^{abc} A_\mu^b J_\mu^c +
k_2 d^{abc} A_\mu^b J_\mu^c = ({\rm c-number}) \one ,}
where $k_{1,2}$ are constants, and
\eqn\efd{f^{abc} = - ~i ~\tr [T^a,T^b] T^c , \quad
d^{abc} = - ~i ~\tr \{T^a, T^b\} T^c .}
The right-hand side of \emotion\
depends on $A_\mu$ and its
derivatives but is irrelevant for the singular coefficient
function and will be considered in sect.~6.  We have obtained
the above form of the equation of motion by demanding
locality;

the equation of motion must be analytic with respect
to the external field $A_\mu$ and its derivatives.  Derivatives
of $A_\mu$ cannot appear on the left-hand side since
the derivatives of $A_\mu$ times the current brings the
scale dimension above 4.

The above equation of motion implies that
\eqn\edC{\eqalign{&\p_\mu C_{\mu\nu}^{ab,\one} (r;A(0))
+ (k_1 f^{acd} + k_2 d^{acd})
A_\mu^c (r) C_{\mu\nu}^{db,\one} (r;A(0))\cr
&\quad\quad \to {\rm finite} \quad {\rm as~} r \to 0 .\cr}}
Substituting \egeneral\ into the above, we
obtain many relations among the constant
coefficients in $D, E, F, G$, and $H$.  The derivation
is straightforward, but tedious.  We give only the results:
\eqn\ek{k_1 = 1,\quad k_2 = 0,}
\eqn\eDrel{D_1 = D_2 \equiv D,\quad D_3 = C+D,\quad D_4 = - 2C-6D,
\quad D_5 = 0,}
\eqn\eFrel{\eqalign{F_4 &=  - {2 \over 3} (\tF_1 + \tF_2) + {4 \over
3} \tF_3 -
{1 \over 15} \tF_5 + {11 \over 15} c_1 \cr
F_6 &= {6 \over 5} (\tF_5 - c_1) \cr
F_7 &= {1 \over 4} (C-D) + 2 \tF_1 + {1 \over 10} \tF_5 - {3 \over 5}
c_1\cr
F_8 &= {1 \over 4} (- C + 3 D) + 2 \tF_2 + {1 \over 10} \tF_5 - {8
\over 5} c_1\cr
F_9 &= F_8 - {D \over 2} + c_1, \quad F_{10} = F_7 - {D \over 2} -
c_1 ,\cr}}
where $c_1$ is a constant, and
\eqn\eFtilde{\eqalign{\tF_1 &\equiv F_1 + F_7,\quad
\tF_2 \equiv F_2 + F_9,\quad \tF_3 \equiv F_3 + F_4 ,\cr
\tF_5 &\equiv F_5 + F_6 + F_7+F_9 ,\cr}}
and
\eqn\eHrel{H_3 = 2 \tH_1 + {1 \over 2} \tH_2 ,}
where
\eqn\eHtilde{\tH_1 \equiv H_1 - {1 \over 2} H_2 + {1 \over 2}
H_3,\quad
\tH_2 \equiv H_2 + H_3 .}
We give relations for $E$ and $G$ in Appendix A.

The equation $k_1=1$ is a choice of the relative
normalization between the gauge field $A_\mu$
and the current $J_\mu$; since the left-hand side
of eq.~\emotion\ is linear in $J_\mu$, but
non-linear in $A_\mu$, we can normalize
$A_\mu$ in any way we like.

But the normalization of the product of $J_\mu$
and $A_\mu$ is fixed by the conjugacy relation
\evev\ (or \evar).  Hence, once we choose $k_1=1$, the
absolute normalization of both $A_\mu$ and $J_\mu$ is fixed.

In order to apply the variational formula \evar\ for $\Phi =
J_\nu^b$,
we need to compute the integral of
\eqn\ediv{\eqalign{&\left[ \d A_\mu^a (r) C_{\mu\nu}^{ab,\one}
(r;A(0))
\right]_{div}\cr
& = (\d A_\mu^a (0)
+ {1 \over 2} r_\alpha r_\beta \p_\alpha \p_\beta \d A_\mu^a (0))
{C \over r^6} \delta^{ab} (\d_{\mu\nu} - 2 \r_\mu \r_\nu) \cr
&\quad+  (\d A_\mu^a (0) + r_\alpha \p_\alpha \d A_\mu^a (0))
{1 \over r^5} D_{\mu\nu}^{ab} (r) \cr
&\quad+ \d A_\mu^a (0) {1 \over r^4} \big(
E_{\mu\nu}^{ab} (r) + F_{\mu\nu}^{ab} (r) +
G_{\mu\nu}^{ab} (r) + H_{\mu\nu}^{ab} (r) \big) .\cr}}
Using eqs.~\egeneral\ -- \eH, \ek\ -- \eHtilde,
and the results in Appendix A, we obtain, after some
calculations, a simple result:
\eqn\esub{\eqalign{
&\int_{1 \ge |r| \ge \ep} {d^4 r \over 2 \pi^2} ~
\left[ \d A_\mu^a (r) C_{\mu\nu}^{ab,\one} (r;A(0)) \right]_{div} \cr
&\quad = {C \over 4} \left( {1 \over \ep^2} -1 \right) \d A_\nu^b (0)
\cr
&\quad\quad - {C \over 12} ~(\ln \ep)~ \tr (-iT^b) \big\lbrace
\p_\nu \p_\alpha A_\alpha - \p^2 A_\nu \cr
&\quad\quad\quad +
[A_\nu, \p_\alpha A_\alpha] + 2 [\p_\alpha A_\nu, A_\alpha]
+ [A_\alpha, \p_\nu A_\alpha] \cr
&\quad\quad\quad + 2 A_\alpha A_\nu A_\alpha - A_\alpha A_\alpha
A_\nu
- A_\nu A_\alpha A_\alpha \big\rbrace \cr
&\quad = {C \over 4} \left( {1 \over \ep^2} -1 \right) \d A_\nu^b (0)
\cr
&\quad\quad - {C \over 12}~ (\ln \ep)~\d~ {\d \over \d A_\nu^b (0)}
\int d^4 r ~{1 \over 4}~\tr \left(\p_\alpha A_\beta - \p_\beta
A_\alpha +
[A_\alpha, A_\beta] \right)^2 .\cr}}

Finally let us write down the variational formula for the correlation
of currents:
\eqn\evarj{\eqalign{&- \d \ave{J_{\nu_1}^{b_1} (r_1)  ...
J_{\nu_n}^{b_n} (r_n)}_A \cr
&\quad = \lim_{\ep \to 0} \Bigg[ \int_{|r-r_i| \ge \ep} d^4 r~
\d A_\mu^a (r) \ave{(J_\mu^a (r) - \ave{J_\mu^a (r)}_A)
J_{\nu_1}^{b_1} (r_1)  ...
J_{\nu_n}^{b_n} (r_n)}_A \cr
&\quad\quad + \sum_{i=1}^n \Big( \d c_{\nu_i}^{b_i,\one}
(A(r_i)) - \int_{1 \ge |r| \ge \ep}
{d^4 r \over 2 \pi^2}
\big[ \d A_\mu^a (r_i + r) C_{\mu\nu}^{ab,\one} (r;A(r_i))
\big]_{div} \Big) \cr
&\qquad\qquad \times \ave{J_{\nu_1}^{b_1} (r_1)  ...
J_{\nu_{i-1}}^{b_{i-1}} (r_{i-1}) J_{\nu_{i+1}}^{b_{i+1}} (r_{i+1})
...
J_{\nu_n}^{b_n} (r_n)}_A \Bigg] .\cr}}
We will examine the finite counterterm $\d c_\nu^{b,\one}(A)$
in the next section.

\newsec{Structure of the connection}

Let us first understand the interpretation of the finite counterterm
$\d c_\mu^{a,\one} (A)$ as a connection.\rqcd\
We can regard the operators
$\{\one, J_\mu^a (r)\}$ as a basis of a vector bundle over
the space of all possible external gauge fields $A_\mu$.
We have seen that a redefinition \echangeF\ induces
a shift of the current $J_\mu^a$ by an identity operator
with an $A_\mu$-dependent coefficient as given by \eredefJ:
\eqn\etrans{\eqalign{\one &\to \one \cr
\tr (- i T^a) J_\mu (r) &\to \tr (-i T^a) J_\mu (r)
+ {\d \over \d A_\mu^a (r)} \int d^4 r' f(A(r')) ~\one  .\cr}}
This is a linear transformation of the basis of the vector bundle.
Under this change of basis, the current-current singularities
are intact.  Hence, the finite counterterm must transform as
\eqn\econn{\d c_\mu^{a,\one} (A(r)) \to
\d c_\mu^{a,\one} (A(r)) = - \d
{\d \over \d A_\mu^a (r)} \int d^4 r' f(A(r'))  .}
This is exactly how a connection transforms, so we can interpret
the finite counterterm as a connection of the vector
bundle.\footnote{$^\dagger$}{The above
discussion needs some refinement;
the operators in the basis $\{\one, J_\mu^a (r)\}$ are not all
independent due to the equation of motion \emotion.}

The connection $\d c_\mu^{a,\one} (A(0))$ must be linear
in $\d A(0)$ and its derivatives, and it contains terms of scale
dimension at most 3.  Hence, it has the structure
\eqn\erough{\d c_\mu^{a,\one} (A(0)) = s_1 \d A_\mu^a (0)
+ ({\rm terms~with~scale~dimension~3}) .}
Terms with scale dimension 2 are absent due to rotational
invariance.
In the following we will narrow down the possible structure of
the connection.

In ref. \rqcd\ we have obtained relations between
finite counterterms and operator singularities by
imposing consistency between variational formulas
and RG equations.  We will now follow the same procedure.
By imposing the consistency between the RG equations
\eRG, \eRGj\ and the variational formula \evarj, we obtain
\eqn\econsist{\d j_\mu^a (A(0)) = \int_{|r|=1} {d^3 r \over 2 \pi^2}~
\left[ \d A_\nu^b (r) C_{\nu\mu}^{ba,\one} (r;A(0)) \right]_{div}
+ \left( 3 - \dl \right) \d c_\mu^{a,\one} (A(0)) ,}
where $j_\mu^a (A(0))$ is defined by \ej.
{}From \esub\ and \erough, we obtain
\eqn\esone{s_1 = - {C \over 4}}
and
\eqn\ejanswer{\eqalign{j_\mu^a (A) &= {C \over 12}~\tr
(-iT^a) \Big(
\p_\mu \p_\alpha A_\alpha - \p^2 A_\mu \cr
&\quad +
[A_\mu, \p_\alpha A_\alpha] + 2 [\p_\alpha A_\mu, A_\alpha]
+ [A_\alpha, \p_\mu A_\alpha] \cr
&\quad + 2 A_\alpha A_\mu A_\alpha - A_\alpha A_\alpha A_\mu
- A_\mu A_\alpha A_\alpha \Big) .\cr}}
This implies
\eqn\eboneanswer{\bone (A) = - {C \over 48} ~\tr
\left( \p_\mu A_\nu - \p_\nu A_\mu + [A_\mu, A_\nu]
\right)^2 .}
Hence, the beta function $\bone (A)$ is $SU(N)$ gauge invariant.

The gauge invariance of $\bone (A)$ is actually expected.
Since the right-hand side of the equation of motion \emotion\
has scale dimension 4, we find
\eqn\eRGmotion{(\dl - 4) D_\mu J_\mu^a = D_\mu j_\mu^a = 0 ,}
where $D_\mu J_\mu^a \equiv \p_\mu J_\mu^a + f^{abc} A_\mu^b
J_\mu^c$.
Hence, $j_\mu^a$ must be a functional derivative of a gauge invariant
integral; $\bone (A)$ must be gauge invariant, up to
a total derivative.  Therefore, we should have expected
\eboneanswer\ up to an overall constant.  Due to eq.~\econsist\
we should have expected the simple answer \esub\
for the integral of the coefficient function as well.

In order to interpret the result \eboneanswer\ more physically,
let us introduce an extra positive parameter $\alpha$ and write the
total
free energy as
\eqn\eFnew{\F[\gone, \alpha, A] = \int d^4 r~\left(
\gone (r) - {1 \over 4 \alpha} \tr F_{\mu\nu}^2 \right) + F[A] ,}
where $F_{\mu\nu} \equiv \p_\mu A_\nu - \p_\nu A_\mu + [A_\mu,
A_\nu]$.
The parameter $\alpha$ plays the role of a gauge coupling
constant, even though the gauge field $A_\mu$ is not dynamical
in our case.  Then the RG equations
\eqn\eRGnew{\eqalign{\dl \gone (r) &= 4 \gone (r) \cr
\dl \alpha &= - {C \over 12}~\alpha^2 \cr
\dl A_\mu &= A_\mu \cr}}
are equivalent to the original RG equations \eRG.
The RG equation of the parameter $\alpha$ agrees with
the  lowest order contribution of the fermion
to the beta function of the gauge coupling constant.

So far we have determined only the dimension 1 term
in the connection, as given by \esone.  To determine
the structure of the dimension 3 terms, we need
further constraint.  We get one constraint by imposing
Maxwell's relation on the second order functional derivative
of the free energy.  We can calculate $F[A + \d_1 A + \d_2 A]$
in two ways:
\eqn\eFmaxwell{\eqalign{F[A + \d_1 A + \d_2 A] &= F[A + \d_1 A] +
\int d^4r~\tr \d_2 A_\mu (r) \ave{J_\mu (r)}_{A+\d_1 A} \cr
&= F[A + \d_2 A] +
\int d^4r~\tr \d_1 A_\mu (r) \ave{J_\mu (r)}_{A+\d_2 A} .\cr}}
By comparing the two using the variational
formula \evarj, we obtain the following commutativity
condition:
\eqn\ecomm{\int d^4 r~\d_1 A_\mu^a (r) \d_2 c_\mu^{a,\one} (A(r))
= \int d^4 r~\d_2 A_\mu^a (r) \d_1 c_\mu^{a,\one} (A(r)) .}
In deriving this result we have used the symmetry
\eqn\esymm{\eqalign{&\int d^4 r~ \d_1 A_\mu^a (r)
\int_{1 \ge |r'| \ge \ep} d^4 r'~\left[\d_2 A_\nu^b (r+r')
C_{\nu\mu}^{ba,\one}
(r';A(r)) \right]_{div} \cr
&\quad = \int d^4 r~ \d_2 A_\mu^a (r)
\int_{1 \ge |r'| \ge \ep} d^4 r'~\left[\d_1 A_\nu^b (r+r')
C_{\nu\mu}^{ba,\one}
(r';A(r)) \right]_{div} ,\cr}}
which is a trivial consequence of eq.~\esub.

The commutativity condition \ecomm\ reduces the number of
possible terms in the connection to a large extent.  Before
enumerating
the possible terms, let us consider one more commutativity condition
following ref.~\rqcd.

The variational formula \evarj\ gives a first order variation
of the correlation function of currents, but we can use it repeatedly
to obtain higher order variations.  We must, then, make sure that
the higher order variations do not depend on the order of taking
variations; in other words we must impose Maxwell's relation
on the second order derivative of the correlation function of
currents.
Thanks to locality of the theory, we need to consider only the
expectation
value of the current.  We can calculate $\ave{J_\gamma^c (0)}_{A+\d_1
A + \d_2 A}$ in two ways:
\eqn\ejmaxwell{\eqalign{&\ave{J_\gamma^c (0)}_{A+\d_1 A+ \d_2 A}\cr
&= \ave{J_\gamma^c (0)}_{A+\d_1 A} \cr
&\quad + \lim_{\ep \to 0} \Bigg[
- \int_{r\ge\ep} d^4 r~ \d_2 A_\mu^a (r) \ave{(J_\mu^a (r) -
\ave{J_\mu^a (r)}_{A+\d_1 A}) J_\gamma^c (0)}_{A+\d_1 A} \cr
&\quad\quad - \d_2 c_\gamma^{c,\one} (A+ \d_1 A(0))
+ \int_{1 \ge |r|\ge\ep} d^4 r~
\left[ \d_2 A_\mu^a (r) C_{\mu\gamma}^{ac,\one} (r;A+\d_1 A(0))
\right]_{div} \Bigg] \cr
&= \ave{J_\gamma^c (0)}_{A+\d_2 A} \cr
&\quad + \lim_{\ep \to 0} \Bigg[
- \int_{r\ge\ep} d^4 r~ \d_1 A_\mu^a (r) \ave{(J_\mu^a (r) -
\ave{J_\mu^a (r)}_{A+\d_2 A}) J_\gamma^c (0)}_{A+\d_2 A} \cr
& \quad\quad - \d_1 c_\gamma^{c,\one} (A+ \d_2 A(0))
+ \int_{1 \ge |r|\ge\ep} d^4 r~
\left[ \d_1 A_\mu^a (r) C_{\mu\gamma}^{ac,\one} (r;A+\d_2 A(0))
\right]_{div} \Bigg] .\cr}}
By using the variational formula \evarj\ once more for each
expression,
we obtain, after some calculations, the consistency condition
\eqn\ecurv{\eqalign{&\d_1 (\d_2 c_\gamma^{c,\one})
- \d_2 (\d_1 c_\gamma^{c,\one})\cr
&\quad =
\int_{1 \ge |r|} d^4 r \int_{1 \ge |r'|} d^4 r'~
\left( - \d_1 A_\mu^a (r) \d_2 A_\nu^b (r')
+ \d_2 A_\mu^a (r) \d_1 A_\nu^b (r') \right) \cr
&\qquad\qquad \times
\ave{J_\mu^a (r) J_\nu^b (r') J_\gamma^c (0)}^c_A ,\cr}}
where $^c$ denotes the connected part.  The left-hand side
gives the curvature of the connection $\d c_\mu^{a,\one}$.
Note the right-hand side is invariant under the redefinition
of the current, given by \etrans.
Therefore, the curvature is invariant under
the gauge transformation of the connection \econn,
as expected.

To summarize our results so far, the dimension 1 term of
the connection is given by \esone, and the dimension 3
terms are constrained by \ecomm\ and \ecurv.  Hence,
we obtain the following most
general form of the connection:
\eqn\egeneralconn{
\eqalign{& \d c_\mu^{a,\one} (A) =
\d {\d \over \d A_\mu^a} \int d^4 r~ \ep (A(r))  - {C \over 4} \d
A_\mu^a \cr
& + \tr i T^a~ \Bigg[
(\Om_1 - \Om_2) (\p_\mu A_\nu \d A_\nu - \p_\nu \d A_\nu A_\mu +
A_\nu \p_\mu \d A_\nu )\cr
&\quad + (- \Om_3 + \Om_4) (\d A_\nu \p_\mu A_\nu - A_\mu \p_\nu \d
A_\nu
+ \p_\mu \d A_\nu A_\nu )\cr
&\quad - \Om_3 (\p_\nu A_\mu \d A_\nu + A_\mu \p_\nu \d A_\nu -
\p_\mu \d A_\nu A_\nu ) \cr
&\quad - \Om_2 (\d A_\nu \p_\nu A_\mu + \p_\nu \d A_\nu A_\mu -
A_\nu \p_\mu \d A_\nu ) \cr
&\quad - \Om_5 (A_\mu A_\nu \d A_\nu + \d A_\nu A_\nu A_\mu) \cr
&\quad - \Om_6 A_\nu \d A_\nu A_\mu - \Om_7 A_\mu \d A_\nu A_\nu \cr
&\quad + \Om_8 ( A_\nu A_\mu \d A_\nu + \d A_\nu A_\mu A_\nu ) \cr
&\quad + {1 \over 2} \Om_9 \d A_\mu \tr A_\nu A_\nu
- \Om_{10} A_\nu \tr \d A_\nu A_\mu - \Om_{11} \d A_\nu
\tr A_\mu A_\nu\cr
& + \ep_{\mu\nu\alpha\beta} \Big\lbrace
t_1 (A_\nu \p_\alpha \d A_\beta + \p_\nu \d A_\alpha A_\beta -
\d A_\nu \p_\alpha A_\beta ) \cr
&\quad + t_2 (\p_\nu A_\alpha \d A_\beta - \d A_\nu \p_\alpha A_\beta
)\cr
&\quad + t_3 (\d A_\nu A_\alpha A_\beta - A_\nu A_\alpha \d
A_\beta)\cr
&\quad + t_4 A_\nu \d A_\alpha A_\beta + t_5 A_\nu \tr A_\alpha \d
A_\beta
\Big\rbrace \Bigg] ,\cr}}
where
\eqn\eepsilon{
\eqalign{\ep (A(r)) &=
a_1 ~\tr \p_\mu A_\nu \p_\mu A_\nu +
a_2 ~\tr \p_\mu A_\mu \p_\nu A_\nu \cr
&
+ a_3 ~\tr \p_\mu A_\nu A_\mu A_\nu +
a_4 ~\tr \p_\mu A_\nu A_\nu A_\mu \cr
&
+ a_5 ~\tr A_\mu A_\mu A_\nu A_\nu +
a_6 ~\tr A_\mu A_\nu A_\mu A_\nu \cr
&
+ a_7 ~\tr A_\mu A_\mu ~\tr A_\nu A_\nu +
a_8 ~\tr A_\mu A_\nu  ~\tr A_\mu A_\nu ,\cr}}
and
the constants $\Om$'s and $t$'s parameterize the curvature
as follows:
\eqn\eomega{
\eqalign{&\d (\d c_\mu^{a,\one}) \cr
& = \tr (iT^a) \Bigg[ \Om_1 \p_\nu \d A_\nu \d A_\mu
 + \Om_2 \d A_\nu \p_\nu \d A_\mu \cr
&\quad - \Om_3 \p_\nu \d A_\mu \d A_\nu -
\Om_4 \d A_\mu \p_\nu \d A_\nu \cr
&\quad + (\Om_1 - \Om_2 - \Om_4) \p_\mu \d A_\nu \d A_\nu \cr
&\quad + (\Om_1 + \Om_3 - \Om_4) \d A_\nu \p_\mu \d A_\nu\cr
&\quad + \Om_5 (\d A_\nu A_\nu \d A_\mu - \d A_\mu A_\nu
\d A_\nu) \cr
&\quad + \Om_6 A_\nu \d A_\nu \d A_\mu
- \Om_7 \d A_\mu \d A_\nu A_\nu \cr
&\quad + (- \Om_5 + \Om_7) A_\mu \d A_\nu \d A_\nu \cr
&\quad + (\Om_5 - \Om_6) \d A_\nu \d A_\nu A_\mu\cr
&\quad + \Om_8 (A_\nu \d A_\mu \d A_\nu -
\d A_\nu \d A_\mu A_\nu)\cr
&\quad - \Om_9  \d A_\mu \tr \d A_\nu A_\nu
- \Om_{10} A_\nu \tr \d A_\mu \d A_\nu\cr
&\quad + \Om_{11} \d A_\nu \tr \d A_\mu A_\nu
+ (\Om_{11} - \Om_{10}) \d A_\nu \tr A_\mu \d A_\nu\cr
&+ \ep_{\mu\nu\alpha\beta} \Big\lbrace
(2 t_1 + t_2) \d A_\nu \p_\alpha \d A_\beta
- (t_1 - t_2) \p_\nu \d A_\alpha \d A_\beta \cr
&\quad - (t_3 - t_4) \d A_\nu \d A_\alpha A_\beta
- 2 t_3 \d A_\nu A_\alpha \d A_\beta \cr
&\quad - (t_3 + t_4) A_\nu \d A_\alpha \d A_\beta\cr
&\quad + t_5 (\d A_\nu \tr A_\alpha \d A_\beta
+ A_\nu \tr \d A_\alpha \d A_\beta) \Big\rbrace \Bigg] .
\cr}}
We have used the notation of differential forms for simplicity.

Note that the $\ep (A(r))$ term in the connection is a pure gauge.
By choosing $f(A(r)) = \ep(A(r))$ in \econn, we can remove, if we
wish,
the pure gauge term by the gauge transformation \econn.
In the next section we will relate the connection \egeneralconn\
to the current-current singularities \egeneral.

\newsec{Determination of the current-current singularities}

So far we have not discussed the right-hand side of the
current equation of motion \emotion:
\eqn\emotionK{
D_\mu J_\mu^a (r) \equiv \p_\mu J_\mu^a (r)  + f^{abc} A_\mu^b (r)
J_\mu^c (r)
= \p_\mu K_\mu^a (A(r)) .}
This is a total derivative due to the global $SU(N)$ invariance.
The possible form of $K_\mu^a (A(r))$ has been studied
extensively, some time ago, regarding its geometrical
significance.\ref\rzumino{B. Zumino, 1983 Les Houches lectures;
W. Bardeen and B. Zumino, Nucl. Phys. {\bf B244}(1984)421}
Let us summarize the known results in the following.
The divergence of the current is related to the $SU(N)$
gauge dependence of the free energy:
\eqn\edF{F[A_\mu - D_\mu \ep] - F[A_\mu]
= \int d^4 r~\tr \ep (r) \p_\mu K_\mu (A(r)) ,}
where $K_\mu = i T^a K_\mu^a, \ep = - i T^a \ep^a$.
The group property of $SU(N)$ implies the Wess-Zumino condition
\ref\rwz{J. Wess and B. Zumino, Phys. Lett. {\bf B37}(1971)95}:
\eqn\ewz{\eqalign{&\int d^4 r~\tr \Big[
\ep (r) \p_\mu (K_\mu (A-D\eta(r)) - K_\mu (A(r)))
- (\ep \leftrightarrow \eta) \Big] \cr
&\quad\quad\quad = \int d^4 r~\tr [\ep, \eta] (r) \p_\mu K_\mu (A(r))
.\cr}}
The most general solution of the Wess-Zumino condition is given by
\eqn\eK{\eqalign{&\int d^4 r~\tr \ep (r) \p_\mu K_\mu (A(r))
= \Delta F[A-D\ep] - \Delta F[A] \cr
&\quad + c \int d^4 r~\tr \ep (r) \ep_{\mu\nu\alpha\beta} \p_\mu
\left(A_\nu\p_\alpha A_\beta + {1 \over 2} A_\nu A_\alpha A_\beta
\right)
,\cr}}
where
\eqn\edeltaF{\eqalign{\Delta F[A] &\equiv
\int d^4 r~ \tr \Big[
g_1 ~\tr \p_\mu A_\nu \p_\mu A_\nu +
g_2 ~\tr \p_\mu A_\mu \p_\nu A_\nu \cr
&
\quad+ g_3 ~\tr \p_\mu A_\nu A_\mu A_\nu +
g_4 ~\tr \p_\mu A_\nu A_\nu A_\mu \cr
&
\quad+ g_5 ~\tr A_\mu A_\mu A_\nu A_\nu +
g_6 ~\tr A_\mu A_\nu A_\mu A_\nu \cr
&
\quad+ g_7 ~\tr A_\mu A_\mu ~\tr A_\nu A_\nu +
g_8 ~\tr A_\mu A_\nu  ~\tr A_\mu A_\nu \Big] .\cr}}

We can remove the $\Delta F$ terms from the equation
of motion by redefining the free energy
$F$ by $F - \Delta F$.  Then, the equation
of motion becomes the familiar anomaly equation \rzumino\
\eqn\emotionsimple{D_\mu J_\mu^a (r) = c ~\p_\mu \cdot
\ep_{\mu\nu\alpha\beta} ~\tr (-iT^a)
\left(A_\nu\p_\alpha A_\beta + {1 \over 2} A_\nu A_\alpha A_\beta
\right) ,}
hence
\eqn\eKsimple{K_\mu^a (A(r)) = c ~\ep_{\mu\nu\alpha\beta} ~\tr
(- i T^a)
\left(A_\nu\p_\alpha A_\beta + {1 \over 2} A_\nu A_\alpha A_\beta
\right) .}
We will adopt this convention from now on.  Then we have
no more freedom to choose $\ep (A(r))$ in \eepsilon\ such as
to remove the pure gauge term in the connection \egeneralconn.

Actually the above choice of the equation of motion
\emotionsimple\ still leaves the current $J_\mu^a$
ambiguous; we can still add an $SU(N)$ gauge
invariant $f(A(r))$ to the free energy.

Under the choice
\eqn\egif{f(A(r)) = {g \over 4} \tr F_{\mu\nu}^2 ,}
the constants $a_1, a_2$ in $\ep (A(r))$, given by \eepsilon,
shift by $2 g, - 2 g$, respectively.  So, by imposing
the gauge condition
\eqn\elastgauge{a_1 = a_2 \equiv a ,}
we will remove the ambiguity of the current entirely.

Now, our task is to assure the consistency of the equation
of motion, given by \emotionsimple\ and \eKsimple, and the
variational
formula \evarj.  We must obtain
\eqn\emotionvar{D_\mu (A+\d A) \ave{J_\mu^a (r)}_{A+\d A} = \p_\mu
K_\mu^a (A+\d A(r))}
from eq.~\emotionsimple\ and the variational formula \evarj.
This requirement gives
\eqn\emotionconsist{\eqalign{
\lim_{\ep \to 0} \Bigg[
& \int_{|r'-r|\ge\ep} d^4 r'~\d A_\nu^b (r') \ave{J_\nu^b (r') \p_\mu
J_\mu^a (r)}^c_A \cr
& - {\p \over \p r_\mu}
\int_{|r'-r|\ge\ep} d^4 r'~\d A_\nu^b (r') \ave{J_\nu^b (r') J_\mu^a
(r)}^c_A
+ f^{abc} \d A_\mu^b (r) \ave{J_\mu^c (r)}_A \cr
& + (\p_\mu \d^{ac} + f^{abc} A_\mu^b (r))
\int_{1\ge |r'| \ge \ep} d^4 r'~
\left[ \d A_\nu^d (r+r') C_{\nu\mu}^{dc,\one} (r';A(r))
\right]_{div} \Bigg] \cr
& = D_\mu \d c_\mu^{a,\one} (A(r)) + \p_\mu \d K_\mu^a (A(r)) .\cr}}

We can evaluate the third integral as follows.  From eq.~\esub\
and \ejanswer, we obtain
\eqn\edivint{\eqalign{I_{2,\mu}^a (r) &\equiv \int_{1\ge |r'| \ge
\ep} d^4 r'~
\left[ \d A_\nu^b (r+r') C_{\nu\mu}^{bc,\one} (r';A(r))
\right]_{div} \cr
& = {C \over 4} \left({1 \over \ep^2} - 1 \right) \d A_\mu^a (r) -
(\ln \ep) \d j_\mu^a (A(r)) .\cr}}
The variation of eq.~\eRGmotion\ gives
\eqn\evardiv{D_\mu \d j_\mu^a + f^{abc} \d A_\mu^b ~j_\mu^c = 0.}
Hence, from \edivint\ and \evardiv, we obtain
\eqn\edivinttwo{D_\mu I_{2,\mu}^a (r) = {C \over 4}
\left({1 \over \ep^2} - 1 \right) D_\mu \d A_\mu^a + (\ln \ep)
f^{abc}
\d A_\mu^b ~j_\mu^c (A(r)) .}

Let us now evaluate the first two integrals in eq.~\emotionconsist.
First we note that we can rewrite them as

\eqn\ediff{\eqalign{
I_1^a (r) &\equiv \int_{|r'-r|\ge\ep} d^4 r'~\d A_\nu^b (r')
\ave{J_\nu^b (r') \p_\mu
J_\mu^a (r)}^c_A \cr
&\quad - {\p \over \p r_\mu}
\int_{|r'-r|\ge\ep} d^4 r'~\d A_\nu^b (r') \ave{J_\nu^b (r') J_\mu^a
(r)}^c_A \cr
&= \int_{|r'-r|=\ep} d^3 r'~{ r'_\mu - r_\mu \over |r'-r|} ~\d
A_\nu^b (r')
\ave{J_\nu^b (r') J_\mu^a (r)}_A^c .\cr}}
Hence, to calculate $I_1^a (r)$ completely, we must know
the current-current singularities to order $1/r^3$.  We now write
\eqn\esingthree{\eqalign{
&2 \pi^2 J_\nu^b (r') J_\mu^a (r) = C_{\nu\mu}^{ba,\one}
(r'-r;A(r)) \one \cr
&\quad + C_{\nu\mu}^{(3)ba,\one} (r'-r;A(r)) \one +
F_{\nu\mu,\alpha}^{ba,c}
(r'-r) J_\alpha^c (r) + o \left({1 \over |r'-r|^3}\right) ,\cr}}
where both $C^{(3)}(r)$ and $F(r)$ are of order $1/r^3$.
We have already studied $C_{\nu\mu}^{ba,\one} (r)$
in sect.~4, so we need to study only $C^{(3)}(r)$ and $F(r)$
for the integral $I_1^a (r)$.

The RG eq. \eRGj\ implies
\eqn\eRGjj{\dl \ave{J_\nu^b (r') J_\mu^a (0)}_A^c =
6 \ave{J_\nu^b (r') J_\mu^a (0)}_A^c .}
Thus, from \esingthree\ and \eRGj, we obtain
\eqn\eRGfc{\eqalign{\dl~ F_{\nu\mu,\alpha}^{ba,c} (r) &=
3~ F_{\nu\mu,\alpha}^{ba,c} (r) \cr
\dl~ C_{\nu\mu}^{(3)ba,\one} (r;A(0)) &= 6  ~C_{\nu\mu}^{(3)ba,\one}
(r;A(0)) -
F_{\nu\mu,\alpha}^{ba,c} (r) ~j_\alpha^c (A(0)) .\cr}}
First we consider $F_{\nu\mu,\alpha}^{ba,c}(r)$.
Using the first of eqs.~\eRGfc, its most general form, satisfying
the symmetry
\eqn\esymmf{F_{\nu\mu,\alpha}^{ba,c}(r) =
F_{\mu\nu,\alpha}^{ab,c}(-r)}
and the equation of motion
\eqn\emotionF{\p_\nu F_{\nu\mu,\alpha}^{ba,c} (r) = 0 ,}
is given by
\eqn\efsol{\eqalign{F_{\nu\mu,\alpha}^{ba,c}(r) = {1 \over r^3}
\Big[ ~&f^{bac} \left( f_1 \r_\nu \r_\mu \r_\alpha + f_2 (\r_\nu
\d_{\mu\alpha}
+ \r_\mu \d_{\nu\alpha} - \r_\alpha \d_{\mu\nu} - 4 \r_\nu \r_\mu
\r_\alpha )
\right) \cr
+~& d^{bac} d_1 ~\ep_{\mu\nu\alpha\beta} \r_\beta~ \Big] .\cr}}
Hence,
\eqn\eintf{\int_{|r|=\ep} {d^3r\over 2 \pi^2}
{}~\r_\mu F_{\nu\mu,\alpha}^{ba,c} (r) =
{f_1 \over 4}~f^{bac} \d_{\nu\alpha} .}
Thus, the integral $I_1^a (r)$ gives the following contribution
proportional to the current:
\eqn\eIonej{I_1^a (r) \to - {f_1 \over 4}~ f^{abc} \d A_\mu^b (r)
\ave{J_\mu^c (r)}_A .}
Therefore, eq.~\emotionconsist\ requires
\eqn\efone{f_1 = 4 .}
Next, we consider the second
of the RG eqs. \eRGfc.  It implies
\eqn\ecthree{
C_{\nu\mu}^{(3)ba,\one} (r;A(0)) =
{1 \over r^3} {\cal A}_{\nu\mu}^{ba}(A(0))
+ (\ln r) F_{\nu\mu,\alpha}^{ba,c} (r) j_\alpha^c (A(0)) ,}
where ${\cal A}(A)$ has scale dimension 3, dependent
only on $A_\mu$ and its derivatives.

Now, from \esingthree, \eIonej, \efone, and \ecthree, we can evaluate
the integral \ediff\ as follows:
\eqn\edifftwo{\eqalign{I_1^a (r) &= \int_{|r'| = \ep} {d^3 r' \over 2
\pi^2}~
\hat{r'}_\mu \d A_\nu^b (r'+r) \Big\lbrace C_{\nu\mu}^{ba,\one}
(r';A(r))
+ {1 \over r'^3} {\cal A}_{\nu\mu}^{ba} (A(r)) \Big\rbrace \cr
&\quad - f^{abc} \d A_\mu^b (r) \ave{J_\mu^c (r)}_A -
(\ln \ep) f^{abc} \d A_\mu^b (r) j_\alpha^c (A(r)) .\cr}}

Therefore, from \edivint\ and \edifftwo, the consistency condition
\emotionconsist\ demands
\eqn\erelation{\eqalign{&\p_\mu \d K_\mu^a (A(r)) +
D_\mu \d c_\mu^{a,\one} (A(r)) =
\lim_{\ep \to 0} \Bigg[
{C \over 4} \left( {1 \over \ep^2} - 1 \right) D_\mu
\d A_\mu^a (r)  \cr
& \quad + \int_{|r'|=\ep} {d^3 r' \over 2 \pi^2}~ \hat{r'}_\mu \d
A_\nu^b (r'+r)
\left\lbrace C_{\nu\mu}^{ba,\one} (r';A(r)) + {1 \over r'^3}
{\cal A}_{\nu\mu}^{ba} (A(r))
\right\rbrace \Bigg] .\cr}}

We can calculate the left-hand side of the above condition
explicitly using \egeneralconn\ and \eKsimple.  The unknown
${\cal A}$ terms cannot give derivatives of $\d A_\mu$.  Hence,
we do not need to know ${\cal A}$
as far as the terms with derivatives of $\d A_\mu$ are concerned.
By calculating only the terms with derivatives of $\d A_\mu$ in
the condition \erelation, we obtain many relations
between the connection \egeneralconn, characterized by the constants
$\Om$'s and $a$'s, and the current-current singularities \egeneral,
characterized by the constants $C$, $D$'s, $E$'s, $F$'s, $G$'s,
and $H$'s.  After some calculations we obtain the following,
which constitutes the main results
of this paper:
\eqn\eathreeafour{\eqalign{a_3 &= {1\over 3} ( 4 \Om_1 - \Om_4) -
{14 \over 3} a \cr
a_4 &= {1 \over 3} (- \Om_1 + 4 \Om_4) + {14 \over 3} a ,\cr}}
where $a$ is given by \elastgauge, and
\eqn\eresult{\eqalign{&C = - 192 a,\quad D = 8 (- \Om_1 + \Om_4 + 8
a)
= 0 \cr
&{1 \over 4} \tF_1 + {1 \over 24} \tF_5 = - {3 \over 2} \Om_1 - \Om_3
+
{3 \over 2} \Om_4 + 8 a\cr
&{1 \over 4} \tF_2 + {1 \over 24} \tF_5 = {3 \over 2} \Om_1 - \Om_2 -
{3 \over 2} \Om_4 - 8 a\cr
&{1 \over 4} \tF_3 + {1 \over 24} \tF_5 =  {1 \over 2} (\Om_1 +
\Om_4)\cr
&c_1 = 0\cr
&{1 \over 4} \tH_1 = - {1 \over 2} t_1 - t_2,\quad
{1 \over 4} \tH_2 = - t_1 + 2 c ,\cr}}
where $c_1$ is defined in \eFrel, and the anomaly constant $c$ is
defined in eq.~\eKsimple.
We give the results for the coefficients $E$'s and $G$'s in
Appendix A.
Eqs.~\eathreeafour\ give two relations among the coefficients
of the connection \egeneralconn.  Eqs.~\eresult\ give
the current-current singularities (but not all of them) in terms
of the connection.
We note that both \eathreeafour\ and \eresult\ are valid irrespective
of the $SU(N)$ representation of the chiral fermion.

In the above derivation we have neglected
the terms of \erelation\ without involving derivatives $\d A_\mu$.
We expect that such terms constrain
the unknown ${\cal A}$ terms but provide no further relations
than what we have obtained above.
Eqs.~\eresult\ are the generalization of the relation between
the finite counterterms and the angular average of the
operator coefficients that we have obtained in ref.~\rqcd.

\newsec{Explicit calculation of the current-current singularities}

In this section we will consider a chiral fermion in the fundamental
representation of $SU(N)$ and compute explicitly both
the current-current singularities
up to first order in $A_\mu$ and the curvature of the connection
at $A_\mu = 0$.  We will then verify the relations
\eresult.

We first consider a free chiral fermion $\psi^i (r) (i=1,...,N)$
in the absence of an external gauge field.    We will suppress
the index $i$ from now on.  The two-point function of the
spinor is given by
\eqn\etwo{\ave{\psi (r) \psibar (0)}_0 = A ~{
1+\gamma_5 \over 2} {\r_\mu \gamma_\mu \over r^3} ,}
where the normalization constant $A$ will be determined later.
The (euclidean) Dirac matrices $\gamma_\mu$ satisfy
\eqn\edirac{\{\gamma_\mu, \gamma_\nu\} = 2 \d_{\mu\nu} .}
We define
\eqn\egammafive{\gamma_5 \equiv \gamma_1
\gamma_2 \gamma_3 \gamma_4}
so that
\eqn\efivetrace{\Sp \gamma_5 \gamma_\mu \gamma_\nu
\gamma_\alpha \gamma_\beta = 4 \ep_{\mu\nu\alpha\beta} ,}
where $\ep_{1234} = 1$.  We have chosen the spinor as
right-handed:
\eqn\ehanded{{1+\gamma_5 \over 2} ~\psi = \psi .}

For $A_\mu = 0$ we can define the current by
\eqn\edefj{J_\mu^a (r) \equiv \psibar (r)
\gamma_\mu T^a \psi (r)  .}
Then, we find the following correlation functions:
\eqn\ejjcor{\ave{J_\mu^a (r) J_\nu^b (0)}_0 = {C \over 2 \pi^2 r^6}
{}~\d^{ab} \left(\d_{\mu\nu} - 2 \r_\mu \r_\nu \right) ,}
where
\eqn\eC{{C \over 2 \pi^2} \equiv - 2 A^2 ,}
and
\eqn\ejjjcor{\ave{J_\mu^a (r') J_\nu^b (r) J_\alpha^c (0)}_0
= {(r'-r)_\beta r_\gamma r'_\d \over r'^4 r^4 (r'-r)^4}~
T_{\mu\nu\alpha,\beta\gamma\d}^{abc} ,}
where
\eqn\eTdef{\eqalign{T_{\mu\nu\alpha,\beta\gamma\d}^{abc}
&\equiv A^3 \Big( \tr T^a T^b T^c ~\Sp \proj~ \gm_\mu
\gm_\beta \gm_\nu \gm_\gm \gm_\alpha \gm_\d\cr
&\quad - \tr T^b T^a T^c ~\Sp \proj~ \gm_\nu \gm_\beta
\gm_\mu \gm_\d \gm_\alpha \gm_\gm \Big) .\cr}}
We note the symmetry
\eqn\eTsymm{T_{\mu\nu\alpha,\beta\gamma\d}^{abc} =
- T_{\nu\mu\alpha,\beta\d\gamma}^{bac} =
- T_{\mu\alpha\nu,\d\gamma\beta}^{acb} .}

We can calculate the current-current singularities by applying the
variational formula \evarj\ to the current-current correlation
function:
\eqn\ejjvar{\eqalign{& - \d \ave{J_\mu^a (r) J_\nu^b (0)}_A\cr
&\quad = \lim_{\ep \to 0} \Bigg[ \int_{|r'| \ge \ep \atop
|r'-r| \ge \ep} d^4 r'~\d A_\alpha^c (r')
\ave{\left( J_\alpha^c (r') - \ave{J_\alpha^c (r')}_A \right)
J_\mu^a (r) J_\nu^b (0)}_A \cr
&\quad\quad + ({\rm counterterms}) \Bigg] .\cr}}
The counterterms do not give any singularity as $r \to 0$.
Hence, by extracting the singular terms of \ejjvar, we obtain
\eqn\eCvar{- \d C_{\mu\nu}^{ab,\one} (r;A(0))
= {\rm Sg}_r \int_{|r'| \ge \ep \atop
|r'-r| \ge \ep} d^4 r'~\d A_\alpha^c (r') \ave{J_\alpha^c (r')
J_\mu^a (r) J_\nu^b (0)}_A^c ,}
where ${\rm Sg}_r$ denotes extracting the part at least as singular
as $1/r^4$.  We can restrict the integral over $r'$ to
any finite radius $r_0 > 0$ without changing the singular terms.
Hence the integral is over a finite region; it is free of infrared
divergences.  (For similar formulas in the case of constant
external fields, see \rphi.)
Taking $A_\mu = 0$ in the above, we obtain
the first order variational formula:
\eqn\eCfirst{- \d C_{\mu\nu}^{ab,\one} (r;A(0))
= {\rm Sg}_r \int_{r_0 \ge |r'| \ge \ep \atop
|r'-r| \ge \ep} d^4 r'~\d A_\alpha^c (r') \ave{J_\alpha^c (r')
J_\mu^a (r) J_\nu^b (0)}_0^c .}
We will compute the right-hand side in the following.

By substituting \ejjjcor\ into \eCfirst, we find
\eqn\estepone{\eqalign{
&- \d C_{\nu\alpha}^{bc,\one} (r;A(0))\cr
&\quad = \d A_\mu^a (0)~
{\rm Sg}_r \int_{r_0 \ge |r'|}
d^4 r'~{(r'-r)_\beta r_\gm r'_\d \over r'^4 r^4 (r'-r)^4}
{}~T_{\mu\nu\alpha,\beta\gm\d}^{abc} \cr
&\quad ~+ \p_\sigma \d A_\mu^a (0)~
{\rm Sg}_r \int_{r_0 \ge |r'|}
d^4 r'~{r'_\sigma (r'-r)_\beta r_\gm r'_\d \over r'^4 r^4 (r'-r)^4}
{}~T_{\mu\nu\alpha,\beta\gm\d}^{abc} .\cr}}
Note that the integrals need no divergent subtraction.
For a particular choice $r_0 = 1$, we find the integrals
as follows:
\eqn\eintegrals{\eqalign{
\int_{1 \ge |r'|} d^4 r'~{(r'-r)_\beta r'_\d \over r'^4 (r'-r)^4} &=
\pi^2 \left[ \left({1 \over 2 r^2} - {1 \over 4} \right)
\d_{\beta \sigma} - {\r_\beta \r_\sigma \over r^2} \right]\cr
\int_{1 \ge |r'|} d^4 r'~{r'_\sigma (r'-r)_\beta r'_\d \over r'^4
(r'-r)^4} &=
{\pi^2 \over r} \Big[
\left({1 \over 4} - {r^2 \over 6} \right)
( \d_{\sigma\beta} \r_\d + \d_{\d\beta} \r_\sigma ) \cr
&\quad + \left( - {1 \over 4} + {r^2 \over 12} \right)
 \d_{\sigma \d} \r_\beta - {1 \over 2}~ \r_\beta \r_\sigma \r_\d
\Big] .\cr}}

By substituting \eintegrals\ into \estepone\ and taking traces
of matrices, we finally obtain
\eqn\eCresult{\eqalign{&2 \pi^2 C_{\nu\alpha}^{bc,\one} (r;A) =
 C \d^{bc} ~{1 \over r^6} (\d_{\nu\alpha} - 2 \r_\nu \r_\alpha)\cr
&\quad -  {8 i \pi^4 A^3 \over r^5}
( \d_{\nu\alpha} - 2 \r_\nu \r_\alpha) \r_\mu \tr [T^b,T^c]
\left(A_\mu + {r_\gm\over 2} \p_\gm A_\mu \right) \cr
&\quad + {2 i \pi^4 A^3 \over r^4} \Big( \tr [T^b,T^c] (\p_\nu
A_\alpha
- \p_\alpha A_\nu) \cr
&\quad\quad - \r_\mu (\r_\nu \tr [T^b, T^c] \p_\mu A_\alpha -
(\nu \leftrightarrow \alpha))
+ \r_\mu (\r_\nu \tr [T^b, T^c] \p_\alpha A_\mu -
(\nu \leftrightarrow \alpha)) \cr
&\quad\quad - (\r_\nu \ep_{\alpha\gm\mu\beta} + \r_\alpha
\ep_{\nu\gm\mu\beta}) \r_\gm \tr \{T^b,T^c\} \p_\mu A_\beta \Big)
+ O (A_\mu^2) .\cr}}
Comparing this result with \eD, we find
\eqn\eDfirst{\eqalign{&D_1 = D_2 = D_5 = 0\cr
&D_3 = - 8 i \pi^4 A^3, \quad D_4 = 16 i \pi^4 A^3 .\cr}}
Hence, from \eDrel\ and \eC, the normalization constant $A$ must be
given by
\eqn\eAresult{A = {1 \over 2 i \pi^2} .}
This is a consequence of the normalization condition \ek, and
assures the spinor field to satisfy the equation of motion
\eqn\espinor{ \gamma_\mu (\p_\mu + i A_\mu^a T^a) \psi (r) = 0 .}
Then the constants $D_3, D_4$ satisfy
the relations given by \eDrel.
Thus, we obtain
\eqn\eCDF{C = {1 \over \pi^2},\quad D =  0 .}
Next, comparing eq.~\eCresult\ with \eF, we find
\eqn\eFfirst{\eqalign{&- F_1 = F_2 = F_7 = - F_8 = - F_9 =
F_{10} = {1 \over 4 \pi^2} \cr
& F_3 = F_4 = F_5 = F_6= 0 ,\cr}}
where we have used \eAresult.
The above gives
\eqn\eFsecond{\tF_1 = \tF_2 = \tF_3 = \tF_5 = c_1 = 0}
and satisfies the relations \eFrel.
Finally, comparing eq.~\eCresult\ with \eH, we find
\eqn\eHfirst{H_1 = 0,\quad H_2 = H_3 = {1 \over 4 \pi^2} .}
This gives \eqn\eHsecond{\tH_1 = 0,\quad \tH_2 = {1 \over 2 \pi^2}}
and satisfies \eHrel.

To verify the relations \eresult\ we must also calculate the
curvature
to determine the constants $\Om$'s and $t$'s through \eomega.
{}From eq.~\ecurv, we obtain
\eqn\ecurvzero{\eqalign{&\d_1 (\d_2 c_\alpha^{c,\one}) -
\d_2 (\d_1 c_\alpha^{c,\one}) \Big|_{A=0}
= \int_{1 \ge |r|} d^4 r \int_{1 \ge |r'|} d^4 r' \cr
&\quad \times (\d_2 A_\mu^a (r) \d_1 A_\nu^b (r') -
\d_1 A_\mu^a (r) \d_2 A_\nu^b (r'))
\ave{J_\mu^a (r) J_\nu^b (r') J_\alpha^c (0)}_0^c .\cr}}
Using the correlation function \ejjjcor, we find
\eqn\ecurvstepone{\eqalign{&\d_1 (\d_2 c_\alpha^{c,\one}) -
\d_2 (\d_1 c_\alpha^{c,\one}) \Big|_{A=0}
= \pi^2 T_{\mu\nu\alpha,\beta\gm\d}^{abc}~\cr
& \quad \times \lim_{\ep \to 0} \Bigg[
(\p_\eta \d_2 A_\mu^a \d_1 A_\nu^b -
\p_\eta \d_1 A_\mu^a \d_2 A_\nu^b) \cr
& \qquad\qquad \times \int_{1 \ge r \ge \ep} d^4 r ~
{r_\d r_\eta \over r^6} \left(
- {1 \over 2} \d_{\beta\gm} + {r_\beta r_\gm \over r^2} +
{r^2 \over 4} \d_{\beta\gm} \right) \cr
& \quad \quad + (\d_2 A_\mu^a \p_\eta \d_1 A_\nu^b -
\d_1 A_\mu^a \p_\eta \d_2 A_\nu^b) \cr
&\qquad\qquad \times \int_{1 \ge r \ge \ep} d^4 r~ {r_\d \over r^6}
\Big\lbrace \left( - {1 \over 4} + {r^2 \over 6} \right)
(\d_{\beta\eta} r_\gm
+ \d_{\beta\gm} r_\eta )\cr
&\qquad\qquad + \left({1 \over 4} - {r^2 \over 12}\right)
\d_{\eta\gm} r_\beta + {1 \over { 2 r^2}} r_\beta r_\gm r_\eta
\Big\rbrace \Bigg] ,\cr}}
where we have used \eintegrals\ for the integrals over $r'$.
In \ecurvstepone\
we have introduced $\ep > 0$ to regulate the divergence of each
integral; the divergences will cancel.

The integrals are obtained as follows:
\eqn\elastintegrals{
\eqalign{
&\int_{1 \ge r \ge \ep} d^4 r ~ {r_\d r_\eta \over r^6} \left(
- {1 \over 2} \d_{\beta\gm} + {r_\beta r_\gm \over r^2} +
{r^2 \over 4} \d_{\beta\gm} \right) \cr
&\quad = 2 \pi^2 \left( - (\ln \ep) ~{1 \over 24} ( \d_{\d\beta}
\d_{\eta\gm}
+ \d_{\d\gm} \d_{\beta\eta} - 2 \d_{\beta\gm} \d_{\eta\d})
+ {1 \over 32} \d_{\beta\gm} \d_{\d\eta} \right) , \cr
&\int_{1 \ge r \ge \ep} d^4 r~ {r_\d \over r^6}
\Bigg[ \left( - {1 \over 4} + {r^2 \over 6} \right) (\d_{\beta\eta}
r_\gm
+ \d_{\beta\gm} r_\eta )\cr
&\qquad\qquad\qquad + \left({1 \over 4} - {r^2 \over 12}\right)
\d_{\eta\gm} r_\beta + {1 \over { 2 r^2}} r_\beta r_\gm r_\eta \Bigg]
\cr
&\quad = 2 \pi^2 \Big(  (\ln \ep)~ {1 \over 24} ( \d_{\beta\gm}
\d_{\d\eta}
+ \d_{\d\gm} \d_{\beta\eta} - 2 \d_{\beta\d} \d_{\eta\gm}) \cr
&\qquad\qquad\qquad + {1 \over 96} (2 \d_{\beta\eta} \d_{\d\gm} +
2 \d_{\beta\gm} \d_{\eta\d} - \d_{\eta\gm} \d_{\beta\d}) \Big) .\cr}}

Hence, substituting the above integrals into
\ecurvstepone\ and using the symmetry \eTsymm, we obtain
\eqn\ecurvsteptwo{\eqalign{&\d_1 (\d_2 c_\alpha^{c,\one}) -
\d_2 (\d_1 c_\alpha^{c,\one}) \Big|_{A=0} \cr
&\quad =
{\pi^4 \over 24} (\p_\eta \d_2 A_\mu^a \d_1 A_\nu^b -
\p_\eta \d_1 A_\mu^a \d_2 A_\nu^b) ~
(T_{\mu\nu\alpha,\eta\beta\beta}^{abc}
+ T_{\mu\nu\alpha,\beta\eta\beta}^{abc}
+ T_{\mu\nu\alpha,\beta\beta\eta}^{abc}) \cr
&\quad = {1 \over 48 \pi^2} ~\tr i T^c \Big(
[\p_\alpha \d_2 A_\mu, \d_1 A_\mu] - (1 \leftrightarrow 2)\cr
&\qquad\qquad\qquad +
[\p_\mu \d_2 A_\mu, \d_1 A_\alpha]  - (1 \leftrightarrow 2)
+ [\p_\mu \d_2 A_\alpha, \d_1 A_\mu]  - (1 \leftrightarrow 2)\cr
&\qquad\qquad\qquad +
3 \ep_{\mu\nu\alpha\eta} \{\p_\eta \d_1 A_\mu , \d_2 A_\nu\}
 - (1 \leftrightarrow 2) \Big),\cr}}
where we have used \eAresult.  Comparing the above result with
\eomega, we find
\eqn\ecurvresult{\eqalign{
 - \Om_1 &= - \Om_2 = \Om_3 = \Om_4 = {1 \over 48 \pi^2} \cr
 t_1 &= - {1 \over 24 \pi^2} ,\quad t_2 = {1 \over 48 \pi^2} .\cr}}

Now, our results are summarized in \eCDF, \eFsecond,
and \eHsecond\ for the operator singularities, and
in \ecurvresult\ for the curvature.  We find that
the relations \eresult\ of the previous section are satisfied, if
we assume
\eqn\eanomaly{c = {1 \over 24 \pi^2} .}
This is the well-known result for the anomaly.  (See, for example,
\rzumino.)
If we take the relations \eresult\ for granted, then we obtain
\eanomaly\ as a consequence of the above calculations.

\newsec{Conclusion}

In this paper we have extended the study of refs.~\rqcd, \rphi\
to the case of space-dependent external fields.

By imposing the validity
of the variational formula \evarj, we have shown that
the connection, given by \egeneralconn, determines most of
the current-current singularities, given by \ejj, \egeneral\ --
\eH, and \ek\ -- \eHtilde, through the relations
\eresult.   We have checked the relations by an explicit
calculation for the chiral fermion in the fundamental representation
of $SU(N)$.

Though we have examined the currents in a non-interacting theory,
it is straightforward to extend our analysis to the flavor currents
in QCD with quarks.  This will be useful for developing perturbation
theory for electroweak interactions, treating QCD exactly; in fact
this is one example discussed in ref.~\rwilson.

In a sense this paper gives mere technical elaboration of the ideas
developed in the classic work of Wilson \rwilson.  Like any good
ideas, they must be fortified by technical arms at some point.

\vfill
\eject
\appendix{A}{Coefficient functions $E(r;A(0)$ and $G(r;A(0)$}

The two remaining coefficient functions $E(r;A(0))$, $G(r;A(0))$
that appear in the OPE \egeneral\ are given as follows:
($\r_\mu \equiv r_\mu/r$)
\eqn\eE{\eqalign{
&E_{\mu\nu}^{ab} (r;A(0)) = E_1 \tr \left( T^a T^b A_\mu A_\nu
+ (a \leftrightarrow b, \mu \leftrightarrow \nu )\right) \cr
&\quad + E_2 \tr \left( T^a T^b A_\nu A_\mu
+ (a \leftrightarrow b, \mu \leftrightarrow \nu )\right) \cr
&\quad + E_3 \tr T^a A_\mu T^b A_\nu + E_4 \tr T^a A_\nu T^b A_\mu
\cr
&\quad + (E_5 \d_{\mu\nu} + E_6 \r_\mu \r_\nu) \tr \{ T^a,T^b\}
A_\alpha A_\alpha
+ (E_7 \d_{\mu\nu} + E_8 \r_\mu \r_\nu) \tr \{T^a, T^b\} \r \cdot A
\r \cdot A \cr
&\quad + (E_9 \d_{\mu\nu} +
E_{10} \r_\mu \r_\nu) \tr T^a A_\alpha T^b A_\alpha
+ (E_{11} \d_{\mu\nu} + E_{12} \r_\mu \r_\nu)
\tr T^a \r \cdot A T^b \r \cdot A \cr
&\quad + E_{13}  \left( \r_\mu \tr T^a T^b \r \cdot A A_\nu
+ (a \leftrightarrow b, \mu \leftrightarrow \nu )\right) \cr
&\quad + E_{14} \left( \r_\nu \tr T^a T^b A_\mu \r \cdot A
+ (a \leftrightarrow b, \mu \leftrightarrow \nu )\right) \cr
&\quad + E_{15} \left( \r_\mu \tr T^a T^b A_\nu \r \cdot A
+ (a \leftrightarrow b, \mu \leftrightarrow \nu )\right) \cr
&\quad + E_{16} \left( \r_\nu \tr T^a T^b \r \cdot A A_\mu
+ (a \leftrightarrow b, \mu \leftrightarrow \nu )\right) \cr
&\quad + E_{17} \left( \r_\mu \tr T^a \r \cdot A T^b A_\nu
+ (a \leftrightarrow b, \mu \leftrightarrow \nu )\right) \cr
&\quad + E_{18} \left( \r_\mu \tr T^a A_\nu T^b \r \cdot A
+ (a \leftrightarrow b, \mu \leftrightarrow \nu )\right) \cr
&\quad + E_{19} \d^{ab} \tr A_\mu A_\nu
+ E_{20} \tr T^a A_\mu \tr T^b A_\nu +
E_{21} \tr T^a A_\nu \tr T^b A_\mu \cr
&\quad + (E_{22} \d_{\mu\nu} + E_{23} \r_\mu \r_\nu) \d^{ab} \tr
A_\alpha A_\alpha
+ (E_{24} \d_{\mu\nu} + E_{25} \r_\mu \r_\nu)
\tr T^a A_\alpha \tr T^b A_\alpha \cr
&\quad + (E_{26} \d_{\mu\nu} + E_{27} \r_\mu \r_\nu) \d^{ab}
\tr \r \cdot A \r \cdot A \cr
&\quad + (E_{28} \d_{\mu\nu} + E_{29} \r_\mu \r_\nu) \tr T^a \r \cdot
A
\tr T^b \r \cdot A \cr
&\quad + E_{30} \d^{ab} \left( \r_\mu \tr \r \cdot A A_\nu
+ (\mu \leftrightarrow \nu )\right) \cr
&\quad + E_{31} \left( \r_\mu \tr T^a \r \cdot A \tr T^b A_\nu
+ (a \leftrightarrow b, \mu \leftrightarrow \nu )\right) \cr
&\quad + E_{32} \left(\r_\mu \tr T^a A_\nu \tr T^b \r \cdot A
+ (a \leftrightarrow b, \mu \leftrightarrow \nu )\right) ,\cr}}
\eqn\eG{\eqalign{&G_{\mu\nu}^{ab} (r;A(0)) = G_1
\ep_{\mu\nu\alpha\beta}
\tr [T^a,T^b] A_\alpha A_\beta\cr
&\quad+ G_2 \ep_{\mu\nu\alpha\beta} \tr T^a A_\alpha T^b A_\beta\cr
&\quad+ G_3 \left(\r_\nu \ep_{\mu\alpha\beta\gamma} \r_\alpha \tr
T^aT^b A_\beta A_\gamma
+ (a \leftrightarrow b, \mu \leftrightarrow \nu )\right) \cr
&\quad+ G_4 \left(\r_\mu \ep_{\nu\alpha\beta\gamma} \r_\alpha \tr
T^a T^b A_\beta A_\gamma
+ (a \leftrightarrow b, \mu \leftrightarrow \nu )\right) \cr
&\quad+ G_5 \left(\r_\nu \ep_{\mu\alpha\beta\gamma} \r_\alpha \tr
T^a A_\beta T^b A_\gamma
+ (a \leftrightarrow b, \mu \leftrightarrow \nu )\right) \cr
&\quad+ G_6 \ep_{\mu\nu\alpha\beta}  \tr T^a A_\alpha \tr T^b A_\beta
\cr
&\quad+ G_7 \left( \r_\nu \ep_{\mu\alpha \beta\gamma} \r_\alpha
\tr T^a A_\beta
\tr T^b A_\gamma
+ (a \leftrightarrow b, \mu \leftrightarrow \nu )\right) . \cr}}

Using \edC\ we obtain the following relations:
\eqn\eErel{\eqalign{E_6&= - {C\over 30} + {2 \over 5} D - {2 \over 3}
(\tE_1+\tE_2)
+ {4 \over 3}\tE_5 - {1 \over 15}\tE_7\cr
E_8&= - {1 \over 5} (2C+6D) + {6 \over 5} \tE_7\cr
E_{10} &= {1 \over 15} C - {4 \over 5} D - {2 \over 3}(\tE_3 + \tE_4)
+ {4 \over 3} \tE_9 - {1 \over 15} \tE_{11}\cr
E_{12} &= {1 \over 5} (4 C + 12 D) + {6 \over 5} \tE_{11}\cr
E_{13} &= {1 \over 10}(3C-D) + 2 \tE_1 +{1 \over 10} \tE_7 + {c_2
\over 2} \cr
E_{15} &= - {C \over 5} - {D \over 10} + 2 \tE_2 + {1 \over 10} \tE_7
- {c_2 \over 2} \cr
E_{14} &= E_{13} + c_2,\quad E_{16} = E_{15} - c_2\cr
E_{17} &= - {C \over 10} + {D \over 5} + 2 \tE_3 + {1 \over 10}
\tE_{11}\cr
E_{18} &= - {C\over 10} + {D \over 5} + 2 \tE_4 + {1 \over 10}
\tE_{11},\cr}}
where $c_2$ is a constant, and
\eqn\eEtilde{\eqalign{
\tE_1 &\equiv E_1 + E_{13},\quad \tE_2 \equiv E_2 + E_{15}\cr
\tE_3 &\equiv E_3 + E_{17},\quad \tE_4 \equiv  E_4 + E_{18}\cr
\tE_5 &\equiv E_5 + E_6,\quad \tE_7 \equiv E_7 + E_8 + E_{13} +
E_{15}\cr
\tE_9 &\equiv E_9 + E_{10},\quad \tE_{11} \equiv E_{11} + E_{12} +
E_{17}
+ E_{18} ,\cr}}
\eqn\eEreltwo{\eqalign{
E_{23} &= - {2 \over 3} \tE_{19} + {4 \over 3} \tE_{22} - {1 \over
15} \tE_{26}\cr
E_{25} &= - {2 \over 3} (\tE_{20}+\tE_{21}) +
{4 \over 3} \tE_{24} - {1 \over 15} \tE_{28} \cr
E_{27} &= {6 \over 5} \tE_{26} ,\quad
E_{29} = {6 \over 5} \tE_{28} \cr
E_{30} &= 2 \tE_{19} + {1 \over 5} \tE_{26} \cr
E_{31} &= 2 \tE_{20} + {1 \over 10} \tE_{28} \cr
E_{32} &= 2 \tE_{22} + {1 \over 10} \tE_{28} ,\cr}}
where
\eqn\eEtildetwo{\eqalign{\tE_{19} &\equiv E_{19} + E_{30} ,\quad
\tE_{20} \equiv E_{20} + E_{31} \cr
\tE_{21} &\equiv E_{21} + E_{32} ,\quad
\tE_{22} \equiv E_{22} + E_{23} \cr
\tE_{24} &\equiv E_{24} + E_{25} ,\quad
\tE_{26} \equiv E_{26} + E_{27} + E_{30} \cr
\tE_{28} &\equiv E_{28} + E_{29} + E_{31} + E_{32} ,\cr}}
and
\eqn\eGrel{\eqalign{G_4 &= - 2 \tG_1 + {1 \over 2} \tG_3 \cr
G_5 &= 2 \tG_2 ,\quad G_7 = 2 \tG_6 ,\cr}}
where
\eqn\eGtilde{\eqalign{\tG_1 &\equiv G_1 + {1 \over 2}(G_3 - G_4)
,\quad
\tG_2 \equiv G_2 + G_5 \cr
\tG_3 &\equiv G_3 + G_4 ,\quad \tG_6 \equiv G_6 + G_7 .\cr}}

{}From the consistency of the equation of motion with
the variational formula we have obtained the condition \erelation.
It gives
\eqn\ectwo{c_2 = 0,}
where $c_2$ is given by \eErel,
\eqn\eGresult{\eqalign{
\tG_1 &= 4 t_3, \quad
{1 \over 4} \tG_2 = 2 t_1 - t_4 + {1 \over 2} c \cr
{1 \over 4} \tG_3 &= - 2 t_1 + c, \quad
\tG_6 = 4 t_5 ,\cr}}
where $t$'s are defined by \egeneralconn, and
\eqn\eEresult{\eqalign{
{1 \over 4} \tE_1 + {1 \over 24} \tE_7 &= {5 \over 3} (\Om_1 - \Om_4)
+ \Om_8
- {28 \over 3} a - 4 a_6 \cr
{1 \over 4} \tE_2 + {1 \over 24} \tE_7 &= {1 \over 3} (- \Om_1 +
\Om_4) -
\Om_5 + {14 \over 3} a - 2 a_5 \cr
{1 \over 4} \tE_3 + {1 \over 24} \tE_{11} &= {4 \over 3} ( - \Om_1 +
\Om_4) -
\Om_7 + {14 \over 3} a - 2 a_5 \cr
{1 \over 4} \tE_4 + {1 \over 24} \tE_{11} &= {4 \over 3} (- \Om_1 +
\Om_4) -
\Om_6 + {14 \over 3} a - 2 a_5 \cr
{1 \over 4} \tE_5 + {1 \over 24} \tE_7 &=  {1 \over 3} (- \Om_1 +
\Om_4) +
{14 \over 3} a - 2 a_5 \cr
{1 \over 4} \tE_9 + {1 \over 24} \tE_{11} &= {2 \over 3} (\Om_1 -
\Om_4) -
{28 \over 3} a - 4 a_6 ,\cr}}
$$
\eqalign{
{1 \over 4} \tE_{19} + {1 \over 12} \tE_{26} &= - \Om_{11} - 4 a_8\cr

{1 \over 4} \tE_{20} + {1 \over 24} \tE_{28} &= -  8 a_7\cr
{1 \over 4} \tE_{21} + {1 \over 24} \tE_{28} &= - \Om_{10} - 4 a_8\cr
{1 \over 4} \tE_{22} + {1 \over 12} \tE_{26} &=  \Om_9 - 8 a_7\cr
{1 \over 4} \tE_{24} + {1 \over 24} \tE_{28} &= - 4 a_8 ,\cr}
$$
where $\Om, a$'s are defined by \eepsilon\ and \egeneralconn.

\appendix{B}{Anomalous commutator}

The derivation of equal-time
commutators from operator product expansions is originally due to
K.~Wilson \rwilson,
and it has been heavily exercised in the study of two dimensional
conformal
field theories more recently.  (See, for example,
\ref\rcft{C. Itzykson et al. (ed.), {\it Conformal invariance
and applications to statistical mechanics} (World Scientific,
1988)}.)

The procedure goes as follows:
First we obtain the
current-current singularities
up to order $1/r^3$ as in \esingthree.  Then we introduce a Wick
rotation
$r_4 \to i r_0$.  Finally, we define
the equal-time commutator as a limit of time-ordered
product of two currents.

In this paper we have discussed the current-current singularities
only
up to order $1/r^4$ in some detail.
Instead of following the above procedure, we will
introduce a euclidean version of a current-current commutator, more
in line with the commutators encountered in conformal field theories.
We will see that for this commutator we need the operator
singularities
only up to order $1/r^4$.

In analogy to the commutators in conformal field theory, we define
the anomalous commutator of two currents by
\eqn\edefcomm{\C_\mu^a (r;\omega) \equiv
\int_{|r'-r|=\ep} d^3r'~ {r'_\nu - r_\nu \over |r' - r|} \omega^b(r')
\ave{J_\nu^b (r')
J_\mu^a (r)}^c_A - f^{abc} \omega^b (r) \ave{J_\mu^c(r)}_A .}
This is anomalous
because of the subtraction of the naive commutator.
Note that the commutator can be divergent as $\ep \to 0$,
corresponding
to divergent Schwinger terms.\rwilson\  For simplicity of notation,
we will consider the commutator at $r=0$ from now on.
We will evaluate the commutator by relating it to the
anomaly equation \emotionsimple; a similar approach has been taken
to evaluate the equal-time commutator of two charge densities
in ref.~\ref\rberry{H. Sonoda, Nucl. Phys. {\bf B266} (1986)410}.

First we recall the following $SU(N)$ gauge dependence of the free
energy:
\eqn\efanomaly{\eqalign{F[A_\mu + D_\mu \omega] &= F[A]
+ \int d^4 r~D_\mu \omega^a (r) \ave{J_\mu^a (r)}_A \cr
& = F[A] - c ~\ep_{\mu\nu\alpha\beta} \int d^4 r~
\tr \omega \p_\mu \left(A_\nu \p_\alpha A_\beta + {1 \over 2}
A_\nu A_\alpha A_\beta \right) ,\cr}}
where $\omega \equiv - i T^a \omega^a$, and
we have used eq.~\evev\ and the anomaly equation \emotionsimple.
By taking the variation of \efanomaly\ with respect to $A_\mu$, we
obtain
\eqn\evarfone{\eqalign{&\d F[A_\mu + D_\mu \omega] = \d F[A] \cr
&\quad - c~ \ep_{\mu\nu\alpha\beta} \int d^4 r~ \tr \d A_\mu
\Big[ \p_\nu A_\alpha \p_\beta \omega + \p_\nu \omega \p_\alpha
A_\beta\cr
&\quad\quad + {1 \over 2} \left( A_\nu A_\alpha \p_\beta \omega
- A_\nu \p_\alpha \omega A_\beta + \p_\nu \omega A_\alpha A_\beta
\right)
\Big] .\cr}}
On the other hand, using eq.~\evev\ directly,
we obtain
\eqn\evarftwo{\d F[A_\mu + D_\mu \omega] = \int d^4 r~ (\d A_\mu^a
+ f^{abc} \omega^b \d A_\mu^c) \ave{J_\mu^a}_{A + D \omega} .}
Hence, from \evarfone\ and \evarftwo, we obtain the $SU(N)$ gauge
dependence of the current:
\eqn\echangej{\eqalign{&\ave{J_\mu^a}_{A+D\omega} - \ave{J_\mu^a}_{A}
= f^{abc} \omega^b \ave{J_\mu^c}_{A} \cr
&\quad +  c~ \ep_{\mu\nu\alpha\beta} \tr (i T^a)
\Big\lbrace \p_\nu A_\alpha \p_\beta \omega
+ \p_\nu \omega \p_\alpha A_\beta\cr
&\quad\quad + {1 \over 2} \left( A_\nu A_\alpha \p_\beta \omega
- A_\nu \p_\alpha \omega A_\beta + \p_\nu \omega A_\alpha A_\beta
\right)
\Big\rbrace .\cr}}
But we can find the gauge dependence also by using the variational
formula \evarj\ directly:
\eqn\echangejtwo{\eqalign{&\ave{J_\mu^a}_{A+D\omega} -
\ave{J_\mu^a}_{A} = - \int_{|r| \ge \ep} d^4 r~ D_\nu \omega^b (r)
\ave{J_\nu^b (r) J_\mu^a (0)}_A^c \cr
&\quad + \int_{1\ge |r|\ge\ep} {d^4 r
\over 2 \pi^2} ~\left[
D_\nu \omega^b (r) C_{\nu\mu}^{ba,\one} (r;A(0))\right]_{div}
- \d_\omega c_\mu^{a,\one} (A(0)) \cr
&= \int_{|r|=\ep} d^3 r~ \r_\nu \omega^b (r) \ave{J_\nu^b (r)
J_\mu^a (0)}_A^c \cr
&\quad + {C \over 4} \left( {1 \over \ep^2} - 1\right) D_\mu \omega^a
(0)
+ \d_\omega j_\mu^a (A(0)) - \d_\omega c_\mu^{a,\one} (A(0)) ,\cr}}
where we have used \esub\ and \ejanswer, $\d_\omega c_\mu^{a,\one}$
is $\d c_\mu^{a,\one}$ with $D \omega$ replacing $\d A$,
and
\eqn\edef{\d_\omega j_\mu^a (A(r)) \equiv j_\mu^a (A+D\omega(r)) -
j_\mu^a (A(r)) .}

Finally, from \echangej\ and \echangejtwo, we obtain the
anomalous commutator:
\eqn\erescomm{\eqalign{\C_\mu^a (0;\omega) = ~
& c~ \ep_{\mu\nu\alpha\beta} ~\tr (i T^a)
\Big\lbrace \p_\nu A_\alpha \p_\beta \omega +
\p_\nu \omega \p_\alpha A_\beta\cr
&\quad\quad + {1 \over 2} \left( A_\nu A_\alpha \p_\beta \omega
- A_\nu \p_\alpha \omega A_\beta + \p_\nu \omega A_\alpha A_\beta
\right)
\Big\rbrace \cr
& + {C \over 4} \left(
1 - {1 \over \ep^2} \right) D_\mu \omega^a (0) -
\d_\omega j_\mu^a (A(0)) + \d_\omega c_\mu^{a,\one} (A(0)) .\cr}}

\listrefs
\parindent=20pt
\bye